# Graph Analysis to Fully Automate Fault Location Identification in Power Distribution Systems

Ali Shakeri Kahnamouei, *Member, IEEE*, and Saeed Lotfifard, *Senior Member, IEEE*

*Abstract*— This paper proposes graph analysis methods to fully automate the fault location identification task in power distribution systems. The proposed methods take basic unordered data from power distribution systems as input, including branch parameters, load values, and the location of measuring devices. The proposed data preparation and analysis methods automatically identify the system's topology and extract essential information, such as faulted paths, structures, loading of laterals and sublaterals, and estimate the fault location accordingly.

The proposed graph analysis methods do not require complex node and branch numbering processes or renumbering following changes in the system topology. The proposed methods eliminate the need for human intervention at any step of the fault location identification process. They are scalable and applicable to systems of any size. The performance of the proposed algorithm is demonstrated using the IEEE 34-bus distribution test system.

## I. INTRODUCTION

FAULT location identification is a critical function in distribution system automation programs. Given its importance, various fault location identification methods have been developed [1]-[10], and several strategies have been proposed in the literature to improve fault location identification results. The advantages and disadvantages of these methods have been discussed in [11]-[12]. However, there is a lack of methodologies in the literature focused on fully automating the data preparation and implementation of fault location methods in distribution systems. For large distribution systems, manually organizing and preparing the input data and equations in the required formats for fault location identification is impractical. This task becomes more challenging when the system undergoes topology changes and reconfigurations, as the input data and equations need to be reorganized each time. Even for a fixed system topology, the faulted path can change depending on the considered fault location. This necessitates human intervention to analyze the system's topology and, based on factors such as the faulted path, the location of measuring devices, and the structure and loading of laterals and sublaterals, manually provide the required information for fault location identification.

Manual data preparation is prone to errors, due to human mistakes in data processing and handling.

Therefore, it is essential to develop a mechanism that can receive basic system information, such as branch parameters and load values, as input. This mechanism should automatically organize and process the data, enabling it to identify the fault location without requiring intervention from system operators.

Graph analysis methods have been proposed to perform various functions in power distribution system analysis and operations. For example, advancements have been made in backward-forward sweep power flow analysis in distribution systems, incorporating graph analysis in some aspects of their methods [13]-[14]. However, to the best of our knowledge, no method has been proposed in the literature that fully automates the task of fault location identification.

The proposed method in this paper has the following key features:

(1) Data preparation and analysis strategies are introduced to fully automate the fault location identification task, eliminating the need for manual processes. The methods take basic system data, such as line parameters and the location of measuring devices, as input. They automatically identify the system's topology, extract the necessary information (such as faulted paths, structures, and loading of laterals and sublaterals), and calculate the fault location.

(2) Branch ordering and path-finding algorithms are proposed that do not require complex node or branch numbering strategies. Additionally, they eliminate the need to renumber nodes when the system topology is reconfigured or when the network is extended.

(3) The input data, which contains network parameter values, typically do not follow any specific order. The proposed method is compatible with such unordered data and does not alter the node numbering, thereby preserving the original numbering. This ensures consistency with the input data, making it easier to verify and debug potential errors.

The remaining of the paper is organized as follows: Section II explains the generic procedures for fault location identification in power distribution systems. Section III explains the proposed data preparation procedures. Section IV explains the automated fault location identification procedure if measuring devices are only installed at the substation. Section V explains the automated fault location identification procedure if in addition to the substation, measuring devices are also installed along the feeder. Section VI represents case study results and finally Section VII provides conclusions.

Shakeri Kahnamouei, and S. Lotfifard are with the School of Electrical Engineering and Computer Science, Washington State University, Pullman, A, 99164, USA, (e-mail: a.shakerikahnamouei@wsu.edu, and s.lotfifard@wsu.edu).

## II. Fault Location Identification Problem

Different fault location methods have been proposed in the literature. However, the generic process can be explained using Fig. 1. Assume a fault occurs at line between nodes (4, 3). The fault location method checks each line of the system one at a time as a hypothetical faulted line to determine if the actual fault has occurred in that line. Starting from the substation node 1, the first hypothetical faulted line is line between nodes (1, 2). Using the measured current and voltages at the substation, the distance of the fault from the upstream node of the line (i.e., node 1) is calculated using the procedure that will be explained later in Section IV. Then, the next hypothetical faulted line is selected which could be line between nodes (2, 5). In this case the voltage and current at node 2 is not known, which should be calculated using the procedure which will be explained later in Sections IV and V. Once the voltages and current at node 2 are calculated the distance of the fault from node 2 on the selected hypothetical faulted line (i.e., line (2, 5)) is determined. This procedure is repeated for all lines of the system one at a time. Finally, the actual fault location is determined using the procedure that will be explained later in Sections IV.C and V.C.

The objective of this paper is to fully automate the above procedure. Algorithm 1 represents the proposed procedures to realize this objective. In the following sections the proposed procedures are explained.

## III. Data preparation process

This section explains the structure of the input data and also describes the proposed data preparation process necessary to perform the automated fault location identification.

### A. Input Data Format

The input data is assumed to consist of basic distribution system information that follows IEEE test systems data format. A radial distribution system's data includes node and branch information. Node data consists of the load values, and branch data includes lines impedance values and possible transformers and regulators information.

The branch information between nodes in a radial distribution system is provided in a matrix, referred to as the B matrix in this paper. It contains N-1 rows, where N is the number of nodes in the system. In the $k^{th}$ row of the B matrix, the first two columns represent the two nodes of the $k^{th}$ branch of the system as an (i,j) pair. The remaining columns of the B matrix provide the values of the branch elements (i.e., the impedance of lines, transformers, and regulators).

For instance, (1) shows the first two columns of the B matrix for the radial network shown in Fig. 1. The first row corresponds to the line between nodes (5, 4).

$$B = \begin{bmatrix} 5 & 4 \\ 7 & 8 \\ 2 & 1 \\ 2 & 5 \\ 8 & 2 \\ 6 & 8 \\ 3 & 4 \\ 2 & 9 \end{bmatrix} \quad (1)$$

The B matrix is typically provided in an unordered form, as the primary goal is to provide the necessary information about the values of the branches in the system (i.e., impedance matrices of the multi-phase line between nodes i and j). In Section III, a procedure is proposed for ordering the B matrix based on the radial configuration of the network.

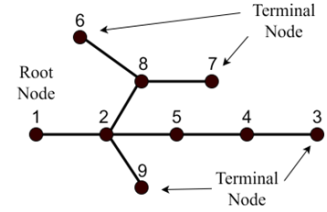

Fig. 1. Schematic of a generic nine-node radial distribution system.

| **Algorithm 1:** The proposed procedures to fully automate the fault location identification task |
|---|
| Step 1: Perform data preparation process using procedures explained in sections III.B, III.C, III.D, and III.E. |
| Step 2: Consider each line of the system as a potential fault location. Apply the procedures described in Step 3 and Step 4 to each line, treating them as a hypothetical faulted line, one at a time. |
| Step 3: If measuring devices are only installed at the substation, calculate the voltages and currents at the upstream terminal node of the selected hypothetical faulted line as described in Section IV.A. Estimate the fault distance for each hypothetical faulted line according to Section IV.B. Finally, determine the actual location of the fault as outlined in Section IV.C. |
| Step 4: If in addition to the substation, measuring devices are also installed along the feeder, calculate the voltages and currents at the upstream terminal node of the selected hypothetical faulted line as described in Section V.A. Then, estimate the fault distance for each hypothetical faulted line according to Section V.B. Finally, determine the actual location of the fault as outlined in Section V.C. |

Additionally, the locations of the measuring devices should be specified. The nodes at which the PMUs (Phasor Measurement Units) and legacy measuring devices are located are stored in vectors called SM (Synchronized Measurements) and LM (Legacy Measurements), respectively.

### B. Presenting Information about the Interconnections of System Nodes in the CN Matrix and NCN Vector

To present the interconnection of nodes in distribution systems, two matrices are defined: Connected Nodes (CN) and Number of Connected Nodes (NCN). Each row in the CN matrix corresponds to a node in the system, and the elements in the $i^{th}$ row represent the nodes that are directly connected to node $i$. Each row of the NCN matrix indicates the total number of nodes connected to the node corresponding to that row. For instance, in Fig. 1 the nodes that are directly connected to node 2 are nodes 1, 5, 8, and 9. Therefore, the second row of CN is (1, 8, 5, 9).

To determine the CN matrix, the following procedure is used: First, an $n \times 1$ empty matrix is defined, which represents $CN = [\ ]_{n \times 1}$. Then, for each branch of the system, which is represented by a pair of (i,j) in B matrix, the element $j$ is inserted into the $j^{th}$ row and $i$ is inserted to the $j^{th}$ row of CN provided they have not already been inserted in that row. By applying this process to all branches in the B matrix (i.e., all rows of B), the connected nodes for each node in the system are determined and stored in the CN matrix.

Additionally, an $n \times 1$ empty matrix is defined, represented as $NCN = [\ ]_{n \times 1}$. The number of elements in each row of the CN matrix is counted and stored in the corresponding row of the NCN matrix. For example, in Fig. 1, the second row of the NCN matrix, which corresponds to node 2, equals 4, as four nodes are connected to node 2.

In Fig.1, the root node of the system (i.e., the substation) and the terminal nodes are shown. As shown in Fig.1, both the root node and the terminal nodes each have one connected node. After determining the NCN vector, the value in the row corresponding to the root node is increased by one. This ensures that the root node is not confused with the terminal nodes, as terminal nodes also have NCN = 1. Therefore, NCN = 1 always refers to the terminal nodes. In Fig. 1, the NCN vector is as follows:

$$NCN = [2 \quad 4 \quad 1 \quad 2 \quad 2 \quad 1 \quad 1 \quad 3 \quad 1]^T \quad (2)$$

In (2) nodes 3, 6, 7, and 9 are terminal nodes as only one node is connected to each of them.

*C. Presenting Branches in an Ordered Manner in OB Matrix*

A matrix with $N - 1$ rows and two columns, referred to as the Ordered Branches (OB) Matrix, stores the branch information of the system in an ordered manner based on the B matrix. The first row of the OB matrix represents one of the last branches of the distribution system, which is connected to the terminal nodes. The row $N - 1$ represents the first branch of the distribution system, which is connected to the root node (i.e., the substation node).

Each row of the OB matrix corresponds to a branch, which is represented by its terminal nodes in an ordered manner. For example, for a branch where the upstream terminal node is node iii and the downstream terminal node is node $j$, the branch is represented in the OB matrix by a pair $(i,j)$, where the upstream node $i$ is stored in the first column and the downstream node $j$ is stored in the second column.

First, all branches from the B matrix are placed in an auxiliary matrix called Aux. Assume the branch in the first row of Aux has terminal nodes labeled $(e, f)$. If the number of connected nodes to node $f$ is one, it indicates that node $f$ does not have any downstream nodes, meaning that the branch $(e, f)$ is one of the last branches in the distribution system. For instance, as shown in Fig. 1, branches like (4,3), (8,7), (8,6), or (2,9) could be the last branches. Therefore, the branch $(e, f)$ is placed in the first row of the OB matrix as $(e, f)$, and it is removed from the Aux matrix. Additionally, in the NCN vector, the value in row $e$ is reduced by one to account for the removal of the connection to node $f$. If the number of connected nodes to node $f$ is not one, node $e$ is then checked. If the number of connected nodes to node $e$ is one, it indicates that node $e$ does not have any downstream nodes either. Therefore, the branch $(e, f)$ is removed from Aux and placed in the first row of the OB matrix as $(f, e)$. If the number of connected nodes to node $e$ and the number of connected nodes to node $f$ are greater than one, it indicates that both nodes have downstream connections. In this case, the next line of the Aux matrix is checked. This process continues until all branches in Aux are sequentially removed and placed in the OB matrix.

The OB matrix for the network in Fig. 1 is shown in (3), which is created by applying the procedure described above to the B matrix in (1).

$$OB = \begin{bmatrix} 8 & 7 \\ 8 & 6 \\ 4 & 3 \\ 5 & 4 \\ 2 & 9 \\ 2 & 5 \\ 2 & 8 \\ 1 & 2 \end{bmatrix} \quad (3)$$

*D. Determining the Paths of the Distribution System*

A path is defined as the route between the root node (i.e., the substation) and a terminal node. It represents the shortest route between two nodes. Since radial distribution systems do not contain loops, there is only one route between any two nodes. Therefore, the number of paths in a radial network is equal to the number of its terminal nodes. In Fig. 1, there are four terminal nodes (i.e., nodes 3, 6, 7 and 9). As a result, there are three paths (Path1: 1-2-5-4-3), (Path 2: 1-2-8-6), (Path 3: 1-2-8-7), and (Path 4: 1-2-9).

The concept of "pathfinding" is widely used in various applications [15] based on different algorithms to find the shortest paths between two nodes, such as Dijkstra's algorithm and the A* algorithm. In this paper, an algorithm is proposed that identifies the paths in a radial distribution system by utilizing the ordered branches of the distribution system (i.e., the OB matrix). This algorithm is straightforward to implement and does not require any renumbering of the nodes.

In the proposed method, a $Paths$ matrix is defined, where each row represents one of the paths in the system, and the non-zero elements of that row indicate the nodes in that path. Initially, the Paths matrix is a $1 \times 1$ empty matrix, which means $Paths\ matrix = [\ ]_{1 \times 1}$. The $Paths$ matrix is populated with branches from the OB matrix until all branches have been incorporated. As previously mentioned, the OB matrix consists of the ordered branches of the distribution system, with the first and last rows representing the last and first branches of the system, respectively.

Assume that the first row of the OB matrix represents a branch with terminal nodes $(nu_1, nd_1)$ where $nu_1$ denotes the node at the upstream side of the first branch and $nd_1$ denotes the node at the downstream side of the first branch. Note that branches in the OB matrix are ordered such that the first column represents the upstream terminal node, and the second column represents the downstream terminal node of the branches.

At the first step, the $Paths$ matrix is initialized as $Paths\ matrix = [nu_1, nd_1\ ]_{1 \times 2}$. Assume that the second row of the OB matrix represents branch $(nu_2, nd_2)$. The branch $(nu_2, nd_2)$ is then checked to determine if it is connected to the branch $(nu_1, nd_1)$. Since $nd_2$ represents the downstream terminal of branch $(nu_2, nd_2)$, if $nd_2 = nu_1$, it means branches $(nu_1, nd_1)$ and $(nu_2, nd_2)$ are connected and branch $(nu_2, nd_2)$ is the upstream branch. In this case, $nu_2$ is inserted into the first column of $Paths$ matrix. Hence, the $Paths$ matrix becomes $Paths\ matrix = [nu_2, nu_1, nd_1\ ]_{1 \times 3}$. If $nd_2 \neq nu_1$, it means that branches $(nu_1, nd_1)$ and $(nu_2, nd_2)$ are not connected, and they belong to different paths. Therefore, a new path is created in the $Paths$ matrix by adding a new row as follows:

$$Paths = \begin{bmatrix} nu_1 & nd_1 \\ nu_2 & nd_2 \end{bmatrix}_{2 \times 2}. \quad (4)$$

For the next selected branch from the OB matrix (i.e., $(nu_3, nd_3)$), if the downstream node of that branch (i.e., $nd_3$) matches the value of any rows of the first column of the $Paths$ matrix, the upstream node of the branch (i.e., $nu_3$) is inserted at the beginning of that row of the $Paths$ matrix. Otherwise, a new path (i.e., a new row) is created in the $Paths$ matrix, and the branch (i.e. $(nu_3, nd_3)$ in this case) is placed in this new row. This procedure is repeated for all branches in the OB.

The $Paths$ matrix for the network in Fig. 1 is as follows:

$$Paths = \begin{bmatrix} 1 & 2 & 8 & 7 & 0 \\ 1 & 2 & 8 & 6 & 0 \\ 1 & 2 & 5 & 4 & 3 \\ 1 & 2 & 9 & 0 & 0 \end{bmatrix}_{4\times 5} \quad (5)$$

*E. Presenting Information about the Structure of Each Node and Line in Matrices*

In this paper, the structure indicates which phases (a, b, and/or c) are included in each node and line of the network. The information about the structure of the lines and nodes should be represented in a format that can be systematically used by the fault location identification algorithm. Two matrices, LineStr and NodeStr, are defined to provide details about whether a given node/line is single-phase, two-phase or three-phase and to specify which phases are involved.

LineStr matrix is a matrix with $N-1$ rows, each representing one of the $N-1$ lines in the system, ordered according to the OB matrix. It has three columns corresponding to phases a, b, and c. For a given line, the corresponding row in the LineStr matrix is considered, and the columns of that row are set to 1 for phases that exist in the line configuration. For non-existing phases, the value is set to 0. For example, if a given line is a two-phase line with phases a and c, the corresponding row in the LineStr matrix would be [1 0 1].

NodeStr matrix is a matrix with N rows, each representing one of the N nodes in the system, and three columns corresponding to phases $a$, b, and c. According to the CN matrix, all nodes connected to a given node are known. Consequently, all lines connected to that node are also known. For a given node, the lines connected to it are identified, and the logical OR of all rows corresponding to these lines in the LineStr matrix is calculated and represented in the NodeStr. For example, in Fig.1, node 4 is connected to nodes 3 and 5. Assume line (5, 4) is a three-phase line, with the corresponding row in the lineStr matrix as [1 1 1]. Also assume line (4, 3) is a two-phase line that includes phases b and c with the corresponding row in the lineStr matrix as [0 1 1]. Therefore, node 4 is a three-phase node, with the corresponding row in the NodeStr matrix being [1 1 1].

## IV. IDENTIFYING THE FAULT LOCATION IF MEASURING DEVICES ARE ONLY INSTALLED AT THE SUBSTATION

The generic procedure for fault location identification was explained in Section II. Conventionally, fault location identification methods rely on measuring devices (voltage and current measurements) installed at the substation. In this section, the procedure for fully automating the fault location identification process is explained, assuming that measuring devices are only installed at the substation. If measuring devices are also installed along the feeder, the procedure explained in Section V should be used

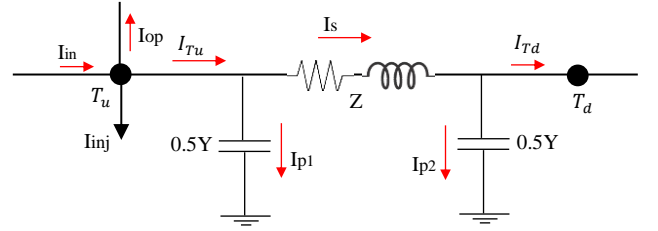

Fig. 2. One line diagram of a distribution system line using $\pi$ model.

*A. Calculation of Voltages and Currents at the Upstream Terminal Node of the Selected Hypothetical Faulted Line*

The voltage and current at the upstream terminal node of the hypothetical faulted line, denoted as $V_u$ and $I_u$, are calculated using the available measurements at the substation.

At the first step, one of the paths containing the hypothetical faulted line is selected from the $Paths$ matrix, and is referred to as the "faulted path". Assume the selected faulted line is line (5, 4) in Fig. 1. Therefore, the third row of the $Paths$ matrix in (5) represents the faulted path, as the hypothetical faulted line (i.e line (5, 4)) lies on that path. As shown in Fig. 2, for each branch, $T_u$ and $T_d$ denote the upstream and downstream terminal nodes of the line, respectively. $I_{Tu}$ and $I_{Td}$ represent the currents at the upstream and downstream terminals of the line, respectively. $I_{in}$ is the input current to node $T_u$, which equals $I_{Td}$ of the previous line on the faulted path. $I_s$ represents the current at the series branch, and $I_{p1}$ and $I_{p2}$ denote the shunt currents of the $\pi$ model of the line. Also, $I_{inj}$ represents the injected current to a node, which is the sum of the load current and DG (distributed generation) current. Additionally, $I_{op}$ at a given node on the faulted path is defined as the sum of the currents of all branches connected to that node that are not part of the faulted path. For example, in Fig. 1, at node 2, $I_{op}$ is the current flowing from node 2 to node 8 and node 9.

Starting from the substation node, the first line on the faulted path is line (1, 2). Therefore, $T_u = 1$ and $T_d = 2$. Also, $V_{Tu}$ and $I_{in}$ are known, as for this line $V_{Tu}$ is the measured voltage at the substation node, and $I_{in}$ is the measured current at the substation node. Next, $I_{op}$ of node $T_u$ should be calculated. For any given node, such as $T_u$ on the faulted path, to calculate $I_{op}$, it must be checked if there are any connected nodes to node $T_u$ that are not on the faulted path. This can be done by checking the corresponding row in the CN vector for node $T_u$. If such a node exists, the current flowing from node $T_u$ to that node must be calculated. This process is carried out using a backward-forward sweep (BFS) method, which is explained in detail later in Section IV.A.1.

Also, the load current at node $T_u$ is calculated as follows [16]:

$$S_L = V_L I_L^* \quad (6)$$

Where $S_L$ denotes the load apparent power.

At the next step, the upstream current of the line (in this case line (1,2)) is calculated as follows:

$$I_{Tu} = I_{in} - I_{op} - I_{inj} \quad (7)$$

Where $I_{inj}$ represents the injected current to the node, which is the sum of the load and DG currents. Next, $I_{p1}$ in Fig.2 is calculated as follows:

$$I_{p1} = YV_{Tu} \quad (8)$$

Then, $I_s$ is obtained as follow:

$$I_s^{pre} = I_{Tu} - I_{p1} \quad (9)$$

$I_s^{pre}$ is a $3 \times 1$ vector representing the current of each phase of the series section of the line. To improve the accuracy of calculation results, the information about the structure of the line (i.e., whether the line is three-phase, two-phase, or single-phase) is incorporated into the calculation of $I_s$ as follows:

$$I_s^{post} = I_s^{pre} \odot lineSTR_i \quad (10)$$

Where $\odot$ is the Hadamard product, representing the element-wise product of vectors.

In (10), if the line is two-phase a-b, for example, the corresponding row of the line structure is [1 1 0]. This ensures that the current in phase c is zero, as the algorithm accounts for the line's configuration.

The voltage of the downstream node and $I_{Td}$ are calculated as follows:

$$V_{Td} = V_{Tu} - ZI_s^{post} \quad (11)$$
$$I_{Td} = I_s^{post} - I_{p2} \quad (12)$$

The above procedure is repeated for the subsequent lines in the faulted path, until the upstream voltage and current of the hypothetical faulted line, i.e., $V_{Tu}$ and $I_{Tu}$, are calculated.

*A.1 Backward-Forward Sweep*

In this section, the process of the BFS method is explained using Fig. 1. As explained in section IV.A, $I_{op}$ must be calculated at nodes that have branched not on the faulted path. For example, for node 2 in Fig.1, $I_{op}$ is the sum of the currents flowing from node 2 to node 8 and from node 2 to node 9. The following steps are followed to calculate $I_{op}$:

***Step 1:*** *Determining the configuration of the subsections connected to the node to which BFS is applied.*

According to CN matrix, the connected nodes to a given node are known. For example, the connected nodes to node 2 are nodes 1, 5, 8, and 9. The nodes on the faulted path are then removed from this set. Thus, the remaining connected nodes are nodes 8 and 9. This means there are two subsections connected to node 2, as branches (2, 8) and (2, 9) are connected to node 2 and are not part of the faulted path. In Fig.1, the subsection associated with branch (2, 8) includes nodes 2, 8, 6, and 7, while the subsection associated with branch (2, 9) consists of nodes 2 and 9. The nodes of each subsection must also be determined automatically. For example, in Fig.1, the nodes downstream of branch (2, 8) are determined as follows: In the *Paths* matrix, all the paths containing line (2, 8) are selected, which are paths (1,2,8,6) and (1,2,8,7). All nodes after node 2 in both paths represent the nodes of the subsection associated with branch (2, 8), which are nodes 2, 8, 6, and 7.

To represent the configuration of a subsection, a matrix called subsection ordered branch (SOB) is defined. Initially, the SOB matrix is set equal to the OB matrix. Then, all the lines in SOB that are represented by nodes 2, 8, 6, and 7 are retained, while the other lines are removed from the SOB matrix. Therefore, the SOB matrix for that subsection containing nodes 2, 8, 7, 6 becomes as follows:

$$SOB = \begin{bmatrix} 8 & 7 \\ 8 & 6 \\ 2 & 8 \end{bmatrix} \quad (13)$$

Note that since the OB matrix is an ordered matrix, the SOB matrix also represents the ordered lines of the subsection.

***Step 2:*** *Initializing the voltages of nodes of subsections.*

The voltages of all the nodes of the subsection are assumed to be equal to the voltage of the root node of the subsection (i.e., node 2 in this example). Since the subsection is radial, all nodes downstream from the root node have either the same phases as the root node or fewer phases. In other words, if the root node is a b-c node, there will be no phase "a" downstream to node 2. Hence, all downstream nodes and lines can have either a b-c structure or a single-phase structure (such as phase b or c). To incorporate this information into the calculation process, after assuming that all node voltages are equal to the root node voltage, each node voltage is multiplied element-wise by the corresponding row in the node structure matrix (NodeStr). This step ensures that the voltages of non-existing phases for any given node are set to zero.

***Step 3:*** *Backward sweep*

In this step, a backward sweep is performed as follows:

Step 3.1: The load currents at each node are calculated using the following equation:

$$I = Z_{load}^{-1} V \quad (14)$$

For each node, this equation is applied only for the existing phases. For non-existing phases, the load current is set to zero. This ensures that the calculation reflects the actual phase configuration at each node.

Step 3.2: Starting from the terminal nodes of the subsection and progressing backward toward the root node, the current flowing from the upstream node is calculated. For example, in Fig. 1, starting from terminal nodes 6 and 7, the current flowing from the upstream node, which is node 8, is calculated. After that, the process moves upstream to calculate the current at the next node in the path, which is node 2. As shown in Fig. 2, $I_s$ is the current flowing over the series reactance of the $\pi$ model of the line. $I_{node}$ at any given node is the sum of $I_{Tu}$ of all downstream branches connected to that node. Initially, $I_{node}$ for all nodes in the subsection is set to zero. The backward process starts from the first row of the SOB matrix. For each line the followings are calculated:

$$I_{node}^{Td} = \sum_i I_{Tu_i} \quad (15)$$
$$I_s = I_{p2} + I_{inj}^{Td} + I_{node}^{Td} \quad (16)$$
$$I_{Tu} = I_{p1} + I_s \quad (17)$$

***Step 4:*** *Forward sweep*

The forward sweep begins by considering the last row of the SOB matrix. This row represents the first branch (or line) in the section. For each branch, the voltage at its downstream terminal node is calculated using the following equation:

$$V_{Td} = V_{Tu} - ZI_s \qquad (18)$$

After calculating the voltage for the downstream terminal node, the voltage vector is element-wise multiplied by the corresponding row in the NodeStr matrix. This ensures that any voltage for non-existing phases is set to zero (ensuring phase consistency). Once all node voltages have been updated, the updated values are compared with the voltages calculated in the previous iteration. If the maximum difference between the updated voltages and the previous iteration's voltages is smaller than a predefined threshold, the BFS process stops. If the difference exceeds the threshold, the next iteration begins from *Step 2*, recalculating voltages and currents.

*B. Estimating the fault distance at each hypothetical faulted line*

Fig. 3 shows a hypothetical faulted line. According to Section IV.A the values of $V_{Tu}$ and $I_{Tu}$ are calculated using the available measurements at the substation. Then, the value of $I_{Tu}$ is multiplied element-wise by the corresponding row of the structure line matrix (LineStr). This step ensures that the current calculation considers phase configuration of the hypothetical faulted line. For example, if the line is a two-phase or single-phase line, the current for the absent phases is set to zero.

The fault distance denoted by d is estimated using the following procedure:

***Step 1:*** *Initialize the fault distance by setting d to 0.5.*
***Step 2:*** *Calculate $I_d$ in Fig. 3 as follows:*
$$I_d = I_{Tu} - 0.5dYV_{Tu} \qquad (19)$$
Then calculate the fault voltage as follows:
$$V_F = V_{Tu} - dZI_d \qquad (20)$$
Where Z and Y are series and shunt values of $\pi$ model of lines

***Step 3:*** *Calculate $I_{d,F}$*

Once $V_F$ is determined in Fig. 3, $I_{d,F}$ can be calculated using the BFS process explained in Section IV. In this case, the root node of the BFS is the faulted point and the system downstream to the faulted node represents the subsystem to which BFS is applied.

***Step 4:*** *Calculate $I_F$*

After calculating $I_{d,F}$, the fault current can be calculated as follows:
$$I_F = I_{u,F} - I_{d,F} \qquad (21)$$

***Step 5:*** *Calculate updated value of d*

For a ground fault, the updated value of d is calculated according to (21).
$$d = \frac{\sum_{\varphi \in \phi} Imag\{V_{Tu}^\varphi (I_F^\varphi)^*\}}{\sum_{\varphi \in \phi} Imag\{V_d^\varphi (I_F^\varphi)^*\}} \qquad (22)$$

For a phase fault, the updated value of d is calculated based on the faulted phases. For example for a phase-a-to-phase-b fault (22) is used.
$$d = \frac{Imag\{(V_{Tu}^b - V_{Tu}^a)(I_F^a)^*\}}{Imag\{(V_d^b - V_d^a)(I_F^a)^*\}} \qquad (23)$$

Where $\varphi$ denotes the set of faulted phases and
$$V_d = ZI_d \qquad (24)$$

***Step 6:*** *Check the stopping criterion*

If the difference between the updated value of d and its previous value is less than a predefined threshold, the process stops. Otherwise, if the updated value of d is between 0 and 1, the process continues to the next iteration starting from *Step 2*. If $d$ is not between 0 and 1, the hypothetical faulted line is considered incorrect, and is removed from the list of possible faulted lines. This indicates that the hypothetical faulted line cannot be the actual faulted line based on the calculated distance. In some cases, especially when there are laterals or sublaterals in the distribution system, multiple lines may have fault distance values $d$ between 0 and 1. This can happen when different lines have the same electrical distances from the substation. In such cases, the algorithm identifies a list of possible faulted lines that could be the actual faulted line.

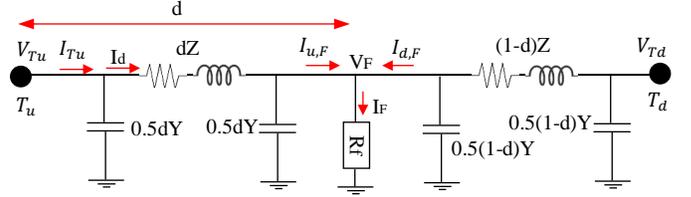

Fig. 3. Schematic of a faulted line.

*C. Eliminate multiple fault location identifications*

If multiple fault locations are identified as the possible location of the fault, further information is needed to eliminate non-faulty lines from the list of identified fault locations. For instance, the method presented in [17] uses information from fault indicators for this purpose. In [18], in addition to fault indicator data, smart meter data is also used.

V. IDENTIFYING THE FAULT LOCATION IN SYSTEMS WITH MEASURING DEVICES INSTALLED ALONG FEEDERS

In modern power distribution systems, measuring devices may also be installed along the feeders. In [19] a method is proposed to utilize data from these measuring devices to improve fault location estimation results. However, it does not provide a procedure for automating the procedures. This section explains how the procedure developed in Section IV can be modified to incorporate these data in calculating $V_{Tu}$, $I_{Tu}$, $V_{Td}$, and $I_{Td}$, thus enhancing the fault location estimation results.

*A. Estimating $V_{Tu}$ and $I_{Tu}$*

The problem of finding $V_{Tu}$ and $I_{Tu}$ is formulated as a state estimation problem, as follows.

***State Variables***: The state variables are defined as:
(1) The real and imaginary parts of the phase currents injected into the nodes of the system.
(2) The real and imaginary parts of each phase of the voltage and current at the upstream node of the hypothetical faulted line, i.e., the real and imaginary parts of $V_{Tu}$ and $I_{Tu}$.

Therefore, the state variables are represented as:
$$x = [V_{Tu}^{i,\phi}, V_{Tu}^{r,\phi}, I_{Tu}^{i,\phi}, I_{Tu}^{r,\phi}, I_n^{i,\phi}, I_n^{r,\phi}] \qquad (25)$$

Where $\phi$ denotes phase a, b, and c. Superscript $i$ denotes the imaginary part, and superscript r denotes real part. $n$ represents the nodes of the system

Note that the system node voltages (i.e. $V_n^{i,\phi}$ and $V_n^{r,\phi}$) are not defined as state variables because they can be expressed as functions of the state variables through the application of Kirchhoff's Current Law (KCL) and Kirchhoff's Voltage Law (KVL) to the network.

***Measurement types*:** The possible measurements in the distribution system can be categorized as follows:
1) Pseudo measurements: Unlike transmission system, not all nodes in a distribution system are equipped with actual measuring devices. Hence, pseudo measurements are used, which include information about the real (P) and reactive (Q) power of the loads. Various methods have been proposed to generate pseudo measurements for load values in distribution systems [20]
2) Actual measurements: These include:
   2.a) Micro PMU measurements: Phasor measurements of node voltages, line currents, and/or load currents.
   2.b) Legacy measurements: Measurements of the magnitude of voltages, the magnitude of line currents, and/or the active and reactive power of the loads.

***Measurement Model*:** To perform state estimation, the equations relating the measurements to the state variables should be defined as follows:
$$h(x) = z \quad (26)$$

(1) Measurement model of pseudo measurements of the loads

For load values, the following relationships hold between the state variables and the pseudo measurements of the loads:
$$V_n^r I_n^r + V_n^i I_n^i = P_n^m \quad (27)$$
$$V_n^i I_n^r - V_n^r I_n^i = Q_n^m \quad (28)$$

The measurement vector is $z = \begin{bmatrix} z_1 \\ z_2 \end{bmatrix} = \begin{bmatrix} P_n^m \\ Q_n^m \end{bmatrix}$, where $P_n^m$ and $Q_n^m$ are the pseudo measurements of active and reactive powers at node n. The measurement model $h(x)$ is defined as $h(x) = \begin{bmatrix} h_1(x) \\ h_2(x) \end{bmatrix} = \begin{bmatrix} V_n^r I_n^r + V_n^i I_n^i \\ V_n^i I_n^r - V_n^r I_n^i \end{bmatrix}$, where $V_n^r$, $V_n^i$, $I_n^r$, and $I_n^i$ are equations expressed as functions of the state variables defined in (25).

(2) Measurement model of *Micro-PMUs*

Micro PMUs may provide synchronized measurements of the phasors of voltages of the nodes, phasors of current injections to the node and/or phasors of line currents. For example, if a micro PMU measures the voltage phasor of node n and the phasor of the line current from node n to node n+1, the following relationships hold between the measurement models and the measurements:
$$V_n^r = V_n^{m,r}, \quad V_n^i = V_n^{m,i} \quad (29)$$
$$I_{n,n+1}^r = I_{n,n+1}^{m,r}, \quad I_{n,n+1}^i = I_{n,n+1}^{m,i} \quad (30)$$

The measurement vector is $z = \begin{bmatrix} z_1 \\ z_2 \\ z_3 \\ z_4 \end{bmatrix} = \begin{bmatrix} V_n^{m,r} \\ V_n^{m,i} \\ I_{n,n+1}^{m,r} \\ I_{n,n+1}^{m,i} \end{bmatrix}$, Where $V_n^{m,r}$ and $V_n^{m,i}$ are the real and imaginary components of the measured voltage phasors at node $n$, while $I_{n,n+1}^{m,r}$ and $I_{n,n+1}^{m,i}$ are the real and imaginary components of the measured line currents from node $n$ toward node $n+1$, respectively. The measurement model, $h(x)$, is also defined as $h(x) = \begin{bmatrix} h_1(x) \\ h_2(x) \\ h_3(x) \\ h_4(x) \end{bmatrix} = \begin{bmatrix} V_n^r \\ V_n^i \\ I_{n,n+1}^r \\ I_{n,n+1}^i \end{bmatrix}$, where $V_n^r$, $V_n^i$, $I_{n,n+1}^r$ and $I_{n,n+1}^i$ are equations expressed as functions of state variables defined in (25)

(3) *Measurement model of legacy measurements*

Legacy measurements only provide information about the magnitude of voltages and currents phasors, as well as active and reactive power values. For instance, if a legacy measuring device measures the voltage magnitude at node n, the magnitude of the phasor of the line current from node n to node n+1, and the active and reactive power from node n toward node n+1, the following relationships hold between the measurement models and the measurements:
$$\sqrt{(V_n^r)^2 + (V_n^i)^2} = |V_n^m| \quad (31)$$
$$\sqrt{(I_{n,n+1}^r)^2 + (I_{n,n+1}^i)^2} = |I_{n,n+1}^m| \quad (32)$$
$$V_n^r I_{n,n+1}^r + V_n^i I_{n,n+1}^i = P_{n,n+1}^m \quad (33)$$
$$V_n^i I_{n,n+1}^r - V_n^r I_{n,n+1}^i = Q_{n,n+1}^m \quad (34)$$

The measurement vector is $z = \begin{bmatrix} z_1 \\ z_2 \\ z_3 \\ z_4 \end{bmatrix} = \begin{bmatrix} |V_n^m| \\ |I_{n,n+1}^m| \\ P_{n,n+1}^m \\ Q_{n,n+1}^m \end{bmatrix}$, where $|V_n^m|$ and $|I_{n,n+1}^m|$ represent the measured voltage magnitude at node n and measured magnitude of line current from node $n$ toward node $n+1$. $P_{n,n+1}^m$ and $Q_{n,n+1}^m$ represent measured active and reactive power from node $n$ toward node $n+1$, respectively.

The measurement model, $h(x)$, is also defined as $\begin{bmatrix} h_1(x) \\ h_2(x) \\ h_3(x) \\ h_4(x) \end{bmatrix} = \begin{bmatrix} \sqrt{(V_n^r)^2 + (V_n^i)^2} \\ \sqrt{(I_{n,n+1}^r)^2 + (I_{n,n+1}^i)^2} \\ V_n^r I_{n,n+1}^r + V_n^i I_{n,n+1}^i \\ V_n^i I_{n,n+1}^r - V_n^r I_{n,n+1}^i \end{bmatrix}$ where $V_n^r$, $V_n^i$, $I_{n,n+1}^r$, and $I_{n,n+1}^i$ are equations expressed as functions of the state variables defined in (25).

***Automatic calculation of $V_{Tu}$ and $I_{Tu}$*:**
First, starting from the upstream terminal node of the hypothetical faulted line and moving backward toward the substation, the measurement models (i.e., h(x) equations) are generated. In this process, state variables are defined as symbolic variables. Node voltages and line currents are calculated as functions of the state variables. This process is similar to the backward sweep process; however, since the state variables are symbolic, a different way of implementation is needed, as follows:

Step 1: The upstream network to the hypothetical faulted line is extracted from the total network OB matrix and is referred to as the upstream ordered branches (UOB) matrix. To create the UOB matrix, the paths containing the hypothetical faulted line are identified, and all the lines downstream of the hypothetical faulted line on those paths are removed from OB and stored in the UOB matrix.

Step 2: Starting from the first row of the UOB matrix, all line and node currents are calculated in a manner similar to the process explained in Section III. However, it is important to note that in this case, all calculations are performed in symbolic form.

Step 3: Go to the first row of the UOB matrix.

Step 4: Check the nodes at both ends of the line. If none of the voltage of the nodes of the line are already presented as a function of the state variables go to step 6, otherwise, if the voltage of any nodes of that line is already written as a function of the state variables, go to step 5.

Step 5: The voltage at the other end of the line is calculated using KVL and the line current which was written as a function of the state variables. Then, update the UOB matrix by removing the row.

Step 6: If there is a next row in UOB matrix, go to the next row of the UOB and go to step 4. Otherwise, go to step 7.

Step 7: If UOB is not empty, go to step 3. Otherwise, stop.

Once the measurement models are developed, the state estimation problem is formulated as follows. To estimate the value of $V_{Tu}$ and $I_{Tu}$ which are among the state variable defined in (25), the following weighted least square state estimation is solved at the $j^{th}$ iteration:

$$x^{j+1} = x^j - [G(x^j)]^{-1} \cdot g(x^j) \quad (35)$$

Where

$$G(x^j) = \frac{\partial g(x^j)}{\partial x} = H^T(x^j) \cdot W \cdot H(x^j) \quad (36)$$

$$g(x^j) = -H^T(x^j) \cdot W \cdot \left(z - h(x^j)\right) \quad (37)$$

$$H(x^j) = \left[\frac{\partial h(x^j)}{\partial x}\right] \quad (38)$$

H(x) is Jacobian matrix of the measurement model, and W is a diagonal weighting matrix with larger weights assigned to more accurate measurements. The iterative process of (36) continues until the convergence criterion $|x^{j+1} - x^j| \leq \varepsilon$ is met.

*B. Estimating the location of the fault at the hypothetical faulted line*

Once the upstream voltage and current of the hypothetical faulted line are estimated as explained above, depending on the availability of measurements at the downstream side of the hypothetical faulted line, the fault location can be calculated as explained in Section IV-C in [21]. In this process, the downstream network to the hypothetical faulted line should be identified automatically using the following procedure.

A faulted path (i.e., a path that contains the hypothetical faulted line) is selected, and all the lines upstream of the hypothetical faulted line on that path are removed from the OB matrix. The remaining lines, which are downstream of the hypothetical faulted line, are then stored in the downstream ordered branches (DOB) matrix. To determine if any measuring devices are installed downstream of the hypothetical faulted line, the nodes in the DOB matrix are checked against the SM and LM vectors.

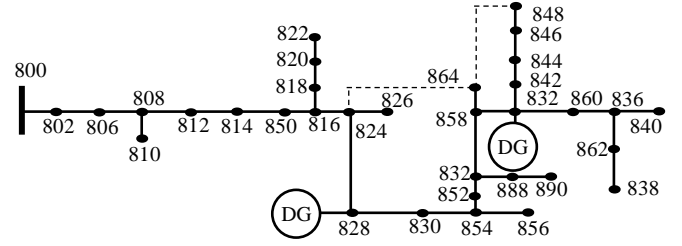

Fig. 4. Modified IEEE 34-bus system

If none of the nodes in the DOB matrix appear in SM or LM vectors, it indicates that no measuring devices are installed downstream of the hypothetical faulted line.

*C. Eliminating possible multiple fault locations*

The procedures explained in Sections V.A and V.B are applied to every line of the system one at a time as the hypothetical faulted line. The lines for which $0 \leq d \leq 1$ are considered in the short list of the possible faulted lines. Finally the actual location of the fault is the one which causes minimum error. The error is defined in [21].

## VI. CASE STUDIES

The proposed data preparation and automated implementation process for fault location identification is demonstrated using the IEEE 34-bus system [22]. The modified IEEE 34-bus system is illustrated in Fig. 4, with two Distributed Generators (DGs) installed at nodes 828 and 832. A micro-PMU and a legacy measuring device are installed at nodes 850 and 858, respectively. Additionally, two tie lines connect nodes 824 to 864 and 864 to 848, which can be utilized for network reconfiguration purposes. The fault location identification process can be applied to any selected configuration. In this case, the process is explained using the main configuration shown in Fig. 4. The structure of the lines in the main configuration is represented in Table I, which indicates the phases (i.e., phases a, b, and c) present at each node and line of the system. The first two columns of Table I, labeled Node 1 and Node 2, represent the B matrix, as discussed in Section III.A. The OB matrix is generated from the B matrix using the procedure explained in Section III.C, and it is shown in Table II. Subsequently, the distribution network paths are identified following the procedure explained in Section III.D. All the paths are illustrated in Fig. 5. Since the proposed branch ordering and path-finding procedure does not require any renumbering of nodes branches, the node labels are retained as in the original IEEE test system data.

After the data preparation process, the fault location identification is performed automatically using the procedure outlined in Section V. Three different fault types (i.e., a-g, a-b-g, and a-b faults) with varying fault resistances (i.e., 0, 10, and 20 ohms) are considered at different locations. The fault location errors for a-g, a-b-g, and a-b faults are presented in Fig. 6, Fig. 7, and Fig. 8, respectively. The error is defined as:

$$Error = \frac{|Estimated\ Location - Actual\ Location|}{Actual\ Location} \quad (39)$$

Where the location of the fault represents the electrical distance from the faulted point to the substation.

TABLE I
NETWORK CONFIGURATION

| Node 1 | Node 2 | Phases | Node 1 | Node 2 | Phases |
|---|---|---|---|---|---|
| 826 | 824 | b | 890 | 888 | a-b-c |
| 818 | 816 | a | 812 | 814 | a-b-c |
| 802 | 800 | a-b-c | 852 | 854 | a-b-c |
| 858 | 832 | a-b-c | 854 | 856 | b |
| 820 | 818 | a | 860 | 836 | a-b-c |
| 816 | 850 | a-b-c | 828 | 824 | a-b-c |
| 858 | 864 | a | 834 | 842 | a-b-c |
| 846 | 848 | a-b-c | 830 | 828 | a-b-c |
| 810 | 808 | b | 808 | 812 | a-b-c |
| 842 | 844 | a-b-c | 858 | 834 | a-b-c |
| 806 | 802 | a-b-c | 854 | 830 | a-b-c |
| 860 | 834 | a-b-c | 832 | 852 | a-b-c |
| 844 | 846 | a-b-c | 836 | 862 | a-b-c |
| 838 | 862 | b | 806 | 808 | a-b-c |
| 832 | 888 | a-b-c | 820 | 822 | a |
| 840 | 836 | a-b-c | 850 | 814 | a-b-c |
| 824 | 816 | a-b-c | | | |

TABLE II
OB MATRIX

| Node 1 | Node 2 | Node 1 | Node 2 | Node 1 | Node 2 |
|---|---|---|---|---|---|
| 800 | 802 | 824 | 826 | 844 | 846 |
| 802 | 806 | 824 | 828 | 846 | 848 |
| 806 | 808 | 828 | 830 | 850 | 816 |
| 808 | 810 | 830 | 854 | 852 | 832 |
| 808 | 812 | 832 | 858 | 854 | 856 |
| 812 | 814 | 832 | 888 | 854 | 852 |
| 814 | 850 | 834 | 860 | 858 | 864 |
| 816 | 818 | 834 | 842 | 858 | 834 |
| 816 | 824 | 836 | 840 | 860 | 836 |
| 818 | 820 | 836 | 862 | 862 | 838 |
| 820 | 822 | 842 | 844 | 888 | 890 |

Fig. 5. Paths of the IEEE 34-bus system.

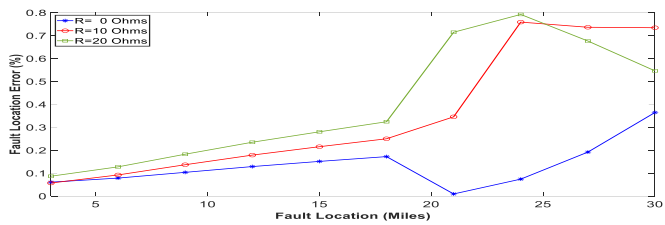
Fig. 6. Estimated fault location error for an a-g fault.

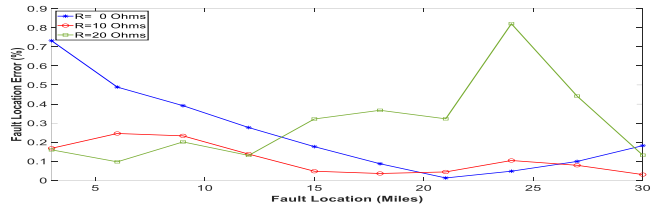
Fig. 7. Estimated fault location error for an a-b-g fault.

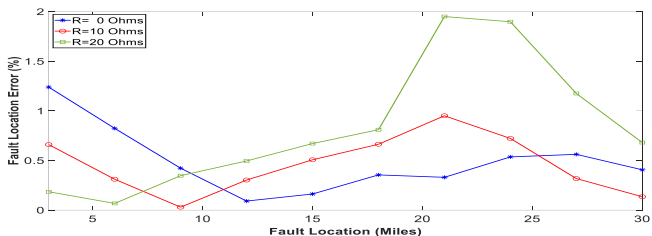
Fig. 8. Estimated fault location error for an a-b fault.

## VII. Conclusions

This paper proposed methods to fully automate the fault location identification process in power distribution systems, aiming to eliminate the need for human intervention. Data preparation procedures were proposed that utilize the system's basic unordered data to automatically identify the system topology and extract the necessary information for fault location identification. Procedures were developed to automatically estimate fault locations in systems with different types of measuring devices. Fully automated fault location process was demonstrated using an IEEE test system.

## References

[1] D. Thukaram, H. P. Khincha, and H. P. Vijaynarasimha, "Artificial neural network and support vector machine approach for locating faults in radial distribution systems", *IEEE Transactions on Power Delivery*, vol. 20, no. 2, pp. 710-721, Apr. 2005.

[2] S. Shi, B. Zhu, A. Lei, X. Dong, "Fault location for radial distribution network via topology and reclosure-generating traveling waves", *IEEE Transactions on Smart Grid*, vol. 10, no. 6, pp. 6404-6413, Nov. 2019.

[3] C. Galvez, and A. Abur "Fault Location in Active Distribution Networks Containing Distributed Energy Resources (DERs)" in *IEEE Transactions on Power Delivery*, vol. 36, no. 5, pp. 3128 – 3139, October 2021.

[4] J. A. Cormane, H. R. Vargas, G. Ordonez, and G. Carrillo "Fault location in distribution systems by means of a statistical model" IEEE Transmission & Distribution Conference and Exposition: Latin America, pp. 1-7, 2006.

[5] S. Lotfifard, M. Kezunovic, and M. J. Mousavi, "Voltage sag data utilization for distribution fault location", *IEEE Transactions on Power Delivery*, vol. 26, no. 2, pp. 1239-1246, Apr. 2011.

[6] R. F. Buzo, H. M. Barradas and F. B. Leão, "A New Method for Fault Location in Distribution Networks Based on Voltage Sag Measurements," in *IEEE Transactions on Power Delivery*, vol. 36, no. 2, pp. 651-662, April 2021.

[7] K. Jia1, Z. Ren, L. Li, Z. Xuan, and D. Thomas "High-frequency transient comparison based fault location in distribution systems with DGs" IET Gener. Transm. Distrib., vol. 11, no. 16, pp. 4068-4077, 2017.

[8] T. A. Short, D. D. Sabin, and M. F. McGranaghan "Using PQ Monitoring and Substation Relays for Fault Location on Distribution Systems" IEEE Rural Electric Power Conferencem, pp.1-7, 2007.

[9] C. G. Arsoniadis, and V. C. Nikolaidis, "Fault Location Method for Overhead Feeders with Distributed Generation Units Based on Direct Load Flow Approach" in *Journal of Modern Power Systems and Clean Energy*, no. 99, pp. 1-11, 2024.

[10] C. A. Apostolopoulos, C. G. Arsoniadis, P. S. Georgilakis, and V. C. Nikolaidis "Unsynchronized Measurements Based Fault Location Algorithm for Active Distribution Systems Without Requiring Source Impedances" in *IEEE Transactions on Power Delivery*, vol. 37, no. 3, pp. 2071 - 2082 June 2022.

[11] J. D. La Cruz, E. Gomez-Luna, M. Ali, J. C. Vasquez, and J. M. Guerrero "Fault Location for Distribution Smart Grids: Literature Overview, Challenges, Solutions, and Future Trends" Energies, vol. 16, pp. 1-37, 2023.

[12] P. Stefanidou-Voziki, N. Sapountzoglou, B. Raison and J.L. Dominguez-Garcia, "A review of fault location and classification methods in distribution grids," in *Electric Power Systems Research*, vol. 209, August 2022.

[13] T. Alinjak, I. Pavic, and M. Stojkov, "Improvement of backward/forward sweep power flow method by using modified breadth-first search strategy," IET Gener. Transmiss. Distrib., vol. 11, no. 1, pp. 102–109, 2017.

[14] S. Ouali and A. Cherkaoui, "An improved backward/forward sweep power flow method based on a new network information organization for radial distribution systems," J. Electr. Comput. Eng., vol. 2020, no. 10, pp. 1–11, 2020.

[15] G. A. Setia, G. HM Sianipar, K. Samudra, F. Haz, N. Winanti and H. R. Iskandar, "Implementation of Backward-Forward Sweep Method on Load Model Variation of Distribution Systems," 2019 2nd International Conference on High Voltage Engineering and Power Systems (ICHVEPS), Denpasar, Indonesia, 2019, pp. 1-5.

[16] D. Foead, A. Ghifari, M. B. Kusuma, N. Hanafiah and E. Gunawan "A Systematic Literature Review of A* Pathfinding" in *Procedia Computer Science*, vol. 179, pp. 508 – 514, 2021.

[17] R. Das, "Determining the locations of faults in distribution systems," Ph.D. dissertation, Elect. Eng. Dept., Saskatchewan Univ., Saskatoon, SK, Canada, 1998.

[18] I. Kiaei, S. Lotfifard, "Fault section identification in smart distribution systems using multi-source data" *IEEE Transactions on Smart Grid*, vol. 11, no.1, pp. 74- 83, 2020.

[19] A. Shakeri Kahnamouei, and S. Lotfifard, "Distribution Systems Fault Location Identification Using Mixed Datasets" *IEEE Transactions on Power Delivery*, early access, 2025.

[20] A. Primadianto and C. N. Lu, "A Review on Distribution System State Estimation," *IEEE Transactions on Power Systems*, vol. 32, no. 5, pp. 3875-3883, Sept. 2017.

[21] A. S. Kahnamouei and S. Lotfifard, "Distribution Systems Fault Location Identification Using Mixed Datasets," in *IEEE Transactions on Power Delivery*, doi: 10.1109/TPWRD.2025.3539734, 2025.

[22] IEEE distribution system analysis subcommittee report, "Radial distribution test feeders," available: https://ewh.ieee.org/soc/pes/dsacom/testfeeders/testfeeders.pdf

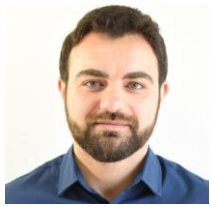

**Ali Shakeri Kahnamouei** (S'19) received his Ph.D. degree from Washington State University, Pullman, WA, USA. His research interests include power system protection and resilience.

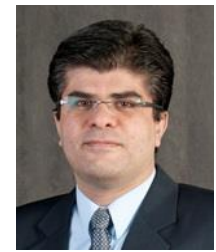

**Saeed Lotfifard** (S'08–M'11-SM'17) received his Ph.D. degree from Texas A&M University. Currently, he is an associate professor at Washington State University, Pullman. His research interests include stability, protection and control of inverter based power systems. He serves as an associate editor for the IEEE Transactions on Power Delivery.